# A novel approach to electron data background treatment in an online wide-angle spectrometer for laser-accelerated ion and electron bunches

F. H. Lindner[1*], J. H. Bin[1,2†], F. Englbrecht[1], D. Haffa[1], P. R. Bolton[1], Y. Gao[1], J. Hartmann[1], P. Hilz[1], C. Kreuzer[1], T. M. Ostermayr[1,3], T. F. Rösch[1], M. Speicher[1], K. Parodi[1], P. G. Thirolf[1], J. Schreiber[1,3‡]

[1]Lehrstuhl für Experimentalphysik – Medizinische Physik, Fakultät für Physik, Ludwig-Maximilians-Universität München, Am Coulombwall 1, 85748 Garching b. München, Germany

[2]Accelerator Technology and Applied Physics Division, Lawrence Berkeley National Laboratory, Berkeley, CA 94720, USA

[3]Max-Planck-Institut für Quantenoptik Garching, Hans-Kopfermann-Str. 1, 85748 Garching b. München, Germany

**Laser-based ion acceleration is driven by electrical fields emerging when target electrons absorb laser energy and consecutively leave the target material. A direct correlation between these electrons and the accelerated ions is thus to be expected and predicted by theoretical models. We report on a modified wide-angle spectrometer allowing the simultaneous characterization of angularly resolved energy distributions of both ions and electrons. Equipped with online pixel detectors, the RadEye1 detectors, the investigation of this correlation gets attainable on a single shot basis. In addition to first insights, we present a novel approach for reliably extracting the primary electron energy distribution from the interfering secondary radiation background. This proves vitally important for quantitative extraction of average electron energies (temperatures) and emitted total charge.**

1. INTRODUCTION

Ion bunches accelerated by the interaction of intense laser pulses with solid targets are distinguished by their very short bunch durations at rather broad energy spreads and their large divergence angles, while they result from a small source. All these properties turn the laser-accelerated ion bunches into a unique ion source, radically differing from conventional acceleration techniques and very promising for a wide range of applications [1-3]. The characterization of typical ion bunch parameters, like the energy spectra and ion bunch divergence, is an indispensable prerequisite for utilizing laser-accelerated ions. Hence, great research efforts are devoted to the development of novel diagnostics.

Thomson parabola spectrometers [4], involving parallelly oriented magnetic and electric fields, are the commonly used devices for ion spectra measurements. However, the typical single pinhole entrance prevents the bunch divergence determination with this detection method. A simultaneous measurement of both energy and angular distribution can be performed using magnetic wide-angle spectrometers [5], providing 1D-angularly resolved energy spectra. So far, ions deflected in such a device are still detected by offline diagnostics like radiochromic films, imaging plates or solid state nuclear track detectors (CR-39). The ongoing advancements in the field of laser-driven ion acceleration, especially with respect to higher repetition rates, however, demand the application of online diagnostics, providing a real-time feedback during laser-plasma experiments. Microchannel plates and scintillators in combination with EMCCD cameras are suitable online detectors, albeit exhibiting substantial drawbacks in terms of costs and time-consuming setups.

---

[*] Electronic mail: florian.lindner@physik.lmu.de
[†] Electronic mail: jianhuibin@lbl.gov
[‡] Electronic mail: joerg.schreiber@lmu.de

In this paper, we present an automated online wide-angle spectrometer (WASP) in combination with large-area pixelated semiconductor detectors [6]. For the first time, it allows a simultaneous and automatic online detection of the energy spectra and angular distribution of both laser-accelerated ions and electrons. We observed a correlation between the measured protons and electrons, as already predicted by various models [7,8] and confirmed in former experiments [9] for the well-known target normal sheath acceleration (TNSA) mechanism [10]. A precise study of this correlation will potentially enable a non-invasive measurement of the ion bunch parameters by detecting the electrons, taking a step towards various applications requiring an unperturbed but well characterized proton or ion bunch.

## 2. EXPERIMENTAL SETUP

The experimental setup of the WASP is shown in Figure 1. Diverging bunches of electrons and ions, generated by the intense laser-plasma interaction, enter the magnetic field of the WASP through a narrow, horizontal slit entrance. The 1D-angular distribution of the particles is measured along the slit axis. The width of the entrance slit limits the energy resolution of the spectrometer. We employed a tooth-like double slit configuration, as sketched in the inset of Figure 1. The first part is a 1.5 mm broad slit made of 2 cm thick steel blocks, thick enough to stop electrons with energies up to 40 MeV [11]. It is followed by a 250 µm broad, tooth-like slit formed by two 1.5 mm thick aluminum plates, providing a sufficiently good energy resolution for the ion detection. This special slit configuration particularly enables the precise determination of the electron signal using a novel method presented in chapter 3.

A fraction of the particle bunch, selected by the slit, enters a magnetic dipole field, where electrons and protons are deflected in opposite directions. The magnets are separated by a

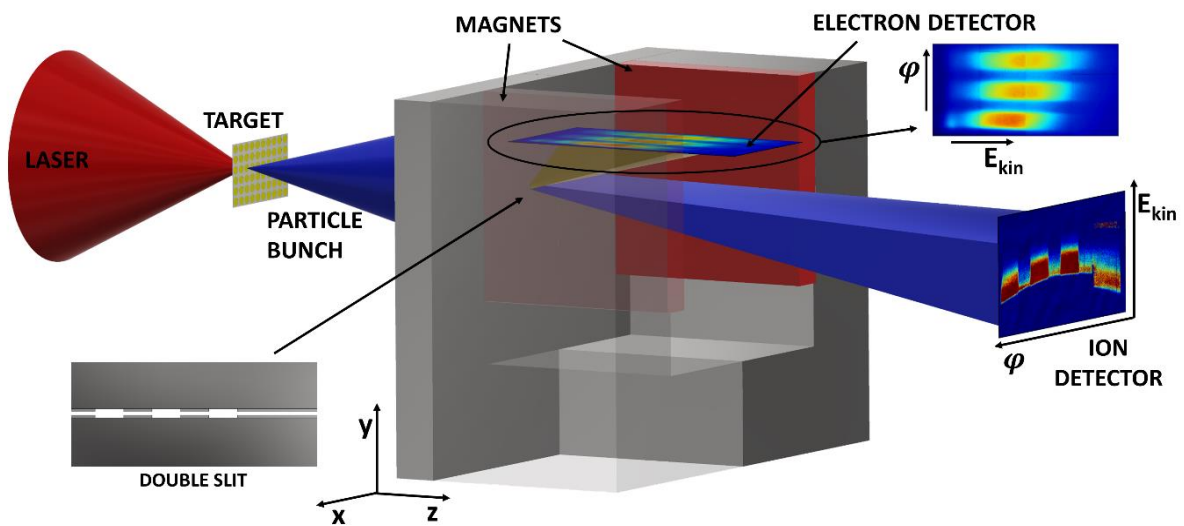

*Figure 1. Schematic sketch of the experimental setup. The laser pulse (red) interacts with a foil target and accelerates a bunch of electrons and ions. These enter a magnetic spectrometer through a thin slit. There, the electrons (yellow) are deflected upwards and detected within the gap between the magnet pole surfaces. The ions (blue) are deflected downwards and detected after an arbitrary dispersion distance. The inset on the left side shows the double slit design of our entrance slit: two 2 cm steel (dark grey), separated by 1.5 mm, followed by a tooth-like array of 250 µm broad aluminum slits (light grey, thickness 1.5 mm).*

105 mm broad gap and provide with a field strength of 150 mT in the center a good compromise in terms of dispersion and energy resolution for both electrons and ions. The magnetic field distribution used for the data analysis was measured with a three-axis Hall magnetometer.

In the present configuration, the ions are deflected downwards and detected behind the magnets. The electrons are deflected upwards and need to be detected already within the gap between the dipole magnets, as their mass is much lower. The amount of deflection of the particles is mainly dependent on their kinetic energy, which thus can be determined by tracing back their trajectories. The angular distribution is given by the lateral dimension, parallel to the magnetic field.

We employed a CMOS-based semiconductor detector, the RadEye1 detector [6,12], as online diagnostics for ions and electrons. One RadEye1 chip consists of 512 x 1024 pixels, each having a size of 48 x 48 µm², thus spanning an active area of 24.6 x 49.2 mm². The 2 µm thick active silicon layer is covered by a silicon oxide passivation layer of the same thickness [13]. Combining four chips, a large detection area of approximately 5 x 10 cm² is achieved with a maximum readout frequency of 2.7 Hz in the current setup. A Kodak BioMax MS Intensifying Screen was directly attached to the surface of the RadEye1 detectors used for electron diagnostics to increase the detection efficiency. An aluminum housing has been designed to protect the electronic detectors from the massive electromagnetic pulse generated as a consequence of the intense laser-matter interaction [14-16]. An aluminum foil with a minimum thickness of 15 µm covers the entrance window in this housing to close the Faraday cage and to protect the detectors from direct and scattered laser light.

## 3. ANGULARLY RESOLVED ENERGY DISTRIBUTION OF PROTONS AND ELECTRONS

The trajectories of the charged particles through the dipole field have been modeled based on the measured magnetic field distribution for the predefined particle energies and angles of incidence. Intersecting these trajectories with the detection plane reveals the respective iso-energy lines along different angles of incidence. We overlaid these iso-energy lines with the unprocessed raw data from the experiment, as indicated in Figure 2(a) for protons. The low energy cut-off curves due to various aluminum foil thicknesses in front of the detector allow a precise calibration of the magnetic field.

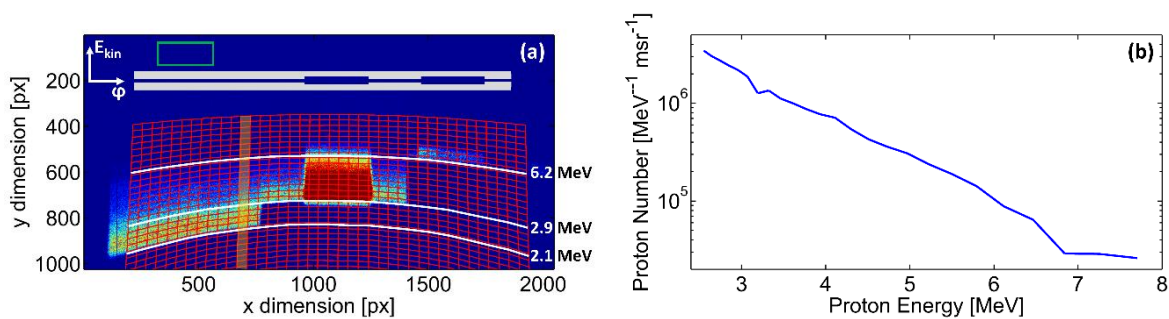

*Figure 2. (a) Proton raw data measured with the RadEye1 detectors, overlaid with the calculated iso-energy, iso-angle map in red. The y axis corresponds to the (energy dependent) magnetic deflection. The x axis is along the spectrometer entrance slit (angle). For the shown example, three differently thick aluminum layers are in front of the RadEye1 detectors in order to calibrate the spectrometer: 45 µm of the left side (protons need a minimum energy of 2.1 MeV to traverse this aluminum layer), 75 µm in the middle (2.9 MeV) and 275 µm on the right side (6.2 MeV). The entrance slit with the two different thicknesses (250 µm and 1.5 mm) is illustrated in the top part of the image. The green rectangle above the sketch of the slit indicates the area used for the determination of the constant mean value for the spectral background determination. (b) Reconstructed exponential proton spectrum at an angle of -0.85° (indicated by the transparent, orange region in the left picture).*

Those iso-energy curves are then used to transfer the two-dimensional spatial information provided by the raw data to the angularly resolved energy distribution of protons (Fig. 2(b)). The pixel values of the RadEye1 detectors are converted into the proton numbers using [6]

$$N_p = \frac{\text{signal}}{\Delta E \cdot 1.09} \frac{\text{ADU}}{\text{keV}}, \qquad (1)$$

where $\Delta E$ denotes the energy loss of a single proton at a given energy within the active layer of a RadEye1 detector pixel, which measures the signal in ADU (analog-to-digital units). This energy loss $\Delta E(E)$ was determined by simulating the energy deposition of the incident protons into the active detection layer of the RadEye1 detector using the software SRIM [11]. The spectral background information is gained by repeating this analysis for an artificially created detector image with each pixel having the same averaged *constant background value*, obtained from an unexposed region within the original raw image, as indicated by the green rectangle in Figure 2(a).

The analysis of the electron data is analogous, but involves more complicated processes. The first step is due to the large angles, at which the electrons impinge at different detector positions. The differences in energy deposition for various propagation distances in the active detection medium must be accounted for. By multiplying the raw data with the sine of the calculated angles between the detection plane and incident particle trajectories, the deposited energy is normalized to the actual thickness of the active layer.

In a second step, we separate the primary electron signal from the background, which is more difficult in case of electron detection compared to protons. This is due to the detector being similarly sensitive for secondary particles and photons generated by the interaction of the incoming primary electrons with the entrance slit and the front plate of the yoked magnet. Therefore, we expect a background that is *not* well represented by a constant offset in each detector pixel, as was assumed in the case of protons.

Instead, we consider the influence of varying widths $d$ of the entrance slit on the electron signal at a position z (along the direction of magnetic deflection and hence energy) on the electron detector. The detected signal $S(z)$ within an area $\Delta z$ along $z(E)$, corresponding to a certain energy interval $\Delta E$ is a transformation of the original signal $S(\vartheta, E)$ incident on the spectrometer entrance slit,

$$S(z) \cdot \Delta z = S(\vartheta, E) \cdot \Delta \vartheta_d \cdot \Delta E, \qquad (2)$$

where $\vartheta$ denotes the angle of particle emission along the y dimension of the spectrometer, and $\Delta \vartheta_d \approx d/L$ is defined by the slit width $d$ and distance $L$ from the source to the spectrometer entrance.

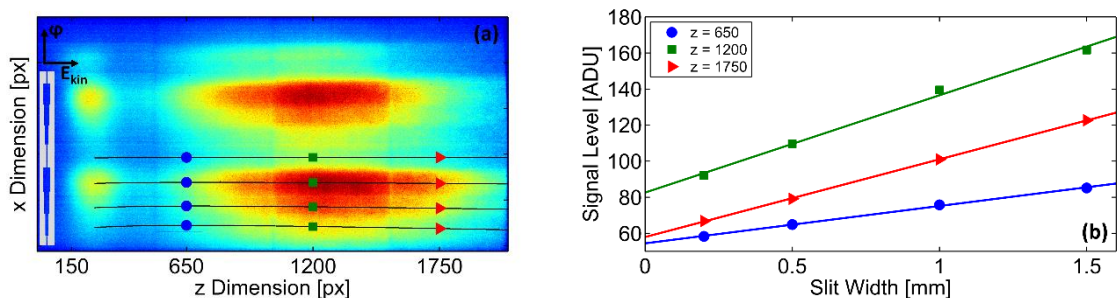

*Figure 3. (a) Electron raw data resulting from a slit configuration with four different slit widths (200 μm, 500 μm, 1 mm and 1.5 mm). The electron bunch enters the spectrometer from the left side, where the entrance slit with the four different widths is illustrated. The x dimension is corresponding to the electron incidence angle (lateral dimension along slit). The z axis is the dimension of magnetic deflection (energy axis). The dotted black lines are iso-angle lines in the center of the slits with varying thickness. The markers denote the positions, at which the data from (b) was taken. (b) Signal levels at different z positions versus the corresponding slit widths. The data points show a linear dependency, as predicted by Equation (2). The data needed for the following calculation of the electron energy spectra, which is done analogous to the protons, can be easily extracted by determining the slopes.*

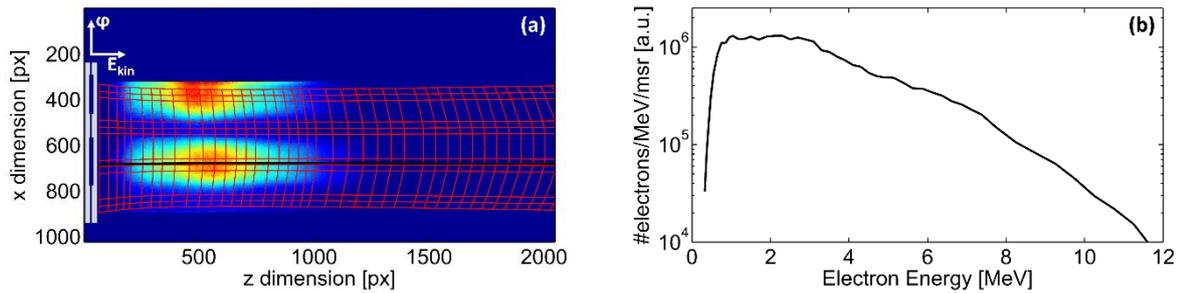

Figure 4. (a) Extracted primary electron data after correction for the angular incidence. The electrons are incident from the left. The slit configuration with two different thicknesses was used, as indicated on the left side of the figure. The x dimension is corresponding to the electrons' incidence angle along the slit, the z axis to the magnetic deflection (energy). The iso-energy, iso-angle map is overlaid with the data (red). The black line indicates the iso-angle line (0.875°), along which the electron energy spectra shown in (b) are extracted from.

Assuming a slowly varying electron signal along the angle of incidence and thus constant electron spectra for adjacent slits with different widths, this linear dependency can be exploited to extract the electron primary signal. Figure 3(a) shows an example of electron data measured in a single shot with a slit configuration providing four different slit widths 0.2, 0.5, 1 and 1.5 mm. As expected from Equation (2), the measured signal levels $S(z)$ depend linearly on the slit width $d$, as demonstrated for three z positions in Figure 3(b). Therefore, the slope of this linear fit is proportional to $S(\vartheta, E) \cdot \frac{\Delta E}{\Delta z}$ and can be readily extracted. Pragmatically, we have reduced the entrance slit for all following measurements to consist of only two different widths instead of the four used to demonstrate this linear behavior.

The electron energy distribution $S(\vartheta, E)$ shown in Figure 4(b) can be determined in the final step using the grid obtained from tracking the electrons through the magnetic fields of the WASP (Figure 4(a)), following the same method as already drafted for protons, and assuming constant energy deposition of primary electrons with energies between 1 – 100 MeV [17,18]. Though not yielding quantitative electron numbers, the signal shape can already serve for drawing fair comparisons between laser shots and determining important electron parameters, for example the mean electron energies (often referred to as electron temperature).

## 4. DISCUSSION AND CONCLUSION

We implemented our WASP as the standard proton and electron diagnostics for all laser-driven ion acceleration experiments using the ATLAS 300 laser at the Laboratory for Extreme Photonics (LEX Photonics) of the LMU Munich in Garching. The 300 TW system delivers laser pulses with energies up to 6 J within pulse durations down to 20 fs at a maximum repetition rate of 5 Hz.

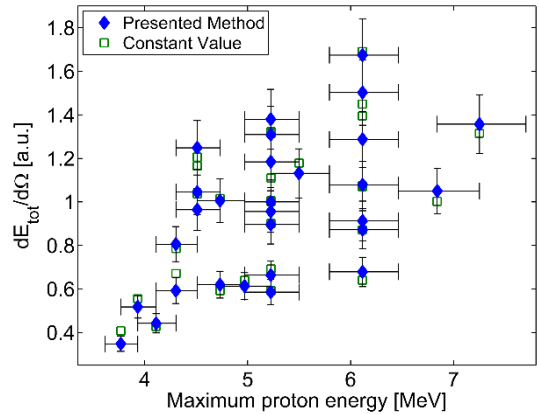

*Figure 5. Differential total electron energies per solid angle, integrated from 1 to 25 MeV, plotted against the maximum proton energies for a number of shots on plastic targets with thicknesses between 250 and 1500 nm. The blue, diamond-shaped data points are evaluated with the new presented method, while the green, rectangular data points are evaluated with the constant background value method; both data series are normalized to an arbitrary value. The error on the x axis is due to the energy resolution of the spectrometer. For the error on the y axis, an error of 20 % was assumed. The errors are only shown for the blue data points, however, they are also directly applicable to the green ones.*

In Figure 5, the total differential electron energy per solid angle, integrated from 1 – 25 MeV, is plotted against the maximum proton energies for a number of shots on plastic targets with thicknesses ranging between 250 – 1500 nm. This observation of a general trend towards higher maximum proton energies with increasing total differential electron energy encourages a closer look into potential electron-proton correlations with higher statistics. This gets attainable by using our novel WASP enabling the simultaneous, single shot online detection of protons and electrons with the possibility of fully exploiting the high repetition rates of the ATLAS laser and comparable systems.

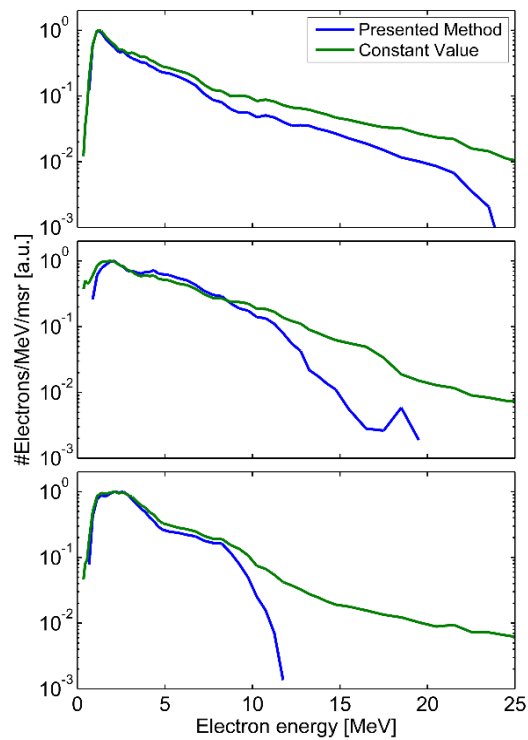

*Figure 6. Comparison of the electron spectra generated with the presented novel method (blue) to the conventional constant background value assumption (green), as also used for the proton analysis, where the background is taken from a not directly irradiated part of the detector. The spectra are normalized to their maxima and shown for three arbitrary laser shots.*

Moreover, the presented approach for the primary electron signal extraction solves formerly hardly accessible issues with the background generated by secondary particles and radiation in electron bunch spectrometry, which always overlays and falsifies the actual electron numbers and the shape of the retrieved energy distribution. To evaluate the importance of our proposed electron data evaluation strategy, we compare it to the constant background value assumption for three arbitrary shots in Figure 6.

As expected, the difference becomes most prominent for large electron energies, where the count rate drops to the noise level. Because this region will decisively influence the determination of the mean electron energy, we judge the novel method for extracting the primary electron signal by introducing the tooth-like slit configuration essential for precise and reliable determination of the detected electron energy spectra in future experiments.

**Acknowledgements**

This work was supported by the BMBF under contract 05P15WMEN9, the DFG Cluster of Excellence MAP (Munich-Centre for Advanced Photonics) and the Centre for Advanced Laser Applications in Garching (CALA). J. H. B. is funded by a Feodor Lynen Fellowship of the Alexander von Humboldt Foundation.